# Design and Testing of Cesium Atomic Concentration Detection System Based on TDLAS

**LZ. Liang,**[a,b,1] **SH. Liu,**[b,a] **ZY. Song,**[c,a] **Y. Wu,**[d,a] **JL. Wei,**[a] **YJ. Xu,**[a] **YH. Xie,**[a] **YL. Xie** [a] and **CD. Hu**[a]

[a] *Hefei Institutes of Physical Science, Chinese Academy of Sciences,*
*Hefei, 230031, China*

[b] *School of Science, Shandong Jianzhu University,*
*Jinan, 250101, China*

[c] *Institute of Physical Science and Information Technology, Anhui University,*
*Hefei, 230601, China*

[d] *School of Electronic and Information Engineering, Anhui Jianzhu University,*
*Hefei, 230601, China*

*E-mail*: lzliang@ipp.ac.cn

ABSTRACT: In order to better build the Neutral Beam Injector with Negative Ion Source (NNBI), the pre-research on key technologies has been carried out for the Comprehensive Research Facility for Fusion Technology (CRAFT). Cesium seeding into negative-ion sources is a prerequisite to obtain the required negative hydrogen ion. The performance of ion source largely depends on the cesium conditions in the source. It is very necessary to quantitatively measure the amount of cesium in the source during the plasma on and off periods (vacuum stage). This article uses the absorption peak of cesium atoms near 852.1nm to build a cesium atom concentration detection system based on Tunable Diode Laser Absorption Spectroscopy (TDLAS) technology. The test experiment based on the cesium cell is carried out, obtained the variation curve of cesium concentration at different temperatures. The experimental results indicate that: the system detection range is within $5 \times 10^6$-$2.5 \times 10^7$ pieces/cm$^3$ and the system resolution better than $1 \times 10^6$ pieces/cm$^3$.

---

[1] Corresponding author.

# Contents

## 1. Introduction

The negative ions in the negative hydrogen ion source are mainly formed by the plasma colliding with hydrogen atoms and positive ions on the low work function surface. The conversion rate depends on the surface work function. In an ion source, cesium deposition on the metal surface reduces the surface work function, thereby increasing the production of negative hydrogen ions in the hydrogen plasma[1]. It is necessary to quantitatively measure the concentration of cesium based on the influence of cesium kinetics on the performance of ion sources[2]. TDLAS technology is currently an advanced gas measurement technology that utilizes the narrow linewidth and adjustable wavelength of tunable semiconductor lasers with injection current to calculate gas concentration by analyzing the selective absorption degree of the laser by the gas to be measured. It has the advantages of high measurement accuracy, strong selectivity, and fast response speed[3].

    This article introduces the design and functional testing of a cesium atom concentration detection system based on TDLAS technology, which detects the concentration and changes of cesium atoms using a cesium vapor pool. Sec.2 will introduce the TDLAS detection principle. Sec.3 will introduce the design principle and structure of the cesium vapor pool. Sec.4 introduces system testing, including qualitative and quantitative detection of cesium.

## 2. TDLAS detection principle

According to Lambert Beer's law, for a single frequency $v$ laser, when passing through an absorbing sample cell, its intensity can be described by the following formula[4]:

$$I(v) = I_0(v)\exp[-\alpha(v)L] = I_0(v)\exp[-\sigma(v)cL] \qquad (2.1)$$

    Where $I$ is the output light intensity, $I_0$ is the incident light intensity, $\alpha(v)$ is the light absorption coefficient of the gas at frequency $v$, $\sigma(v)$ is the light Absorption cross section of the gas at frequency $v$, $L$ is the optical path length of the sample cell, and $c$ is the concentration of the absorption gas.

In order to reduce the interference of noise signals, wavelength modulation technology is usually used to modulate the output wavelength of the laser. When the laser with the center frequency of $v_e$ is modulated by the modulation wave with the frequency of $\omega$, its Instantaneous phase and frequency can be expressed as:

$$v = v_e + \delta_v \cos \omega t \qquad (2.2)$$

Where, $\delta_v$ is the modulation amplitude, and the intensity of light passing through the sample cell can be expressed by the cosine Fourier series of $I(v_e)$:

$$I(v_e, t) = \sum_{n=0}^{\infty} A_n(v_e) \cos(n\omega t) \qquad (2.3)$$

Each harmonic component $A_n$ of the harmonic can be measured through the Lock-in amplifier:

$$A_n(v_e) = \frac{2I_0 cL}{\pi} \int_0^\pi -\sigma(v_e + \delta_v \cos\theta) \cos n\theta\, d\theta \qquad (2.4)$$

In the above equation, $\theta = \omega t$. Therefore, each harmonic component is directly positively correlated with gas concentration, and second harmonic is usually used for detection in experiments.

## 3. Design of cesium vapor pool

### 3.1 Design principle and structure of cesium cell

The system uses a controllable cesium vapor pool for experimental research. The design principle of controllable cesium vapor pool is based on saturated Vapor pressure and Ideal gas law. Under closed conditions, the pressure of vapor in phase equilibrium with solids or liquids at a certain temperature is called saturated Vapor pressure. The same substance has different saturated Vapor pressure at different temperatures, and increases with the increase of temperature. Because the Ideal gas law is an ideal model established by people to simplify the Real gas, the higher the temperature, the lower the pressure, and the closer to the Ideal gas. In practice, the relevant ideal conditions cannot be realized, so the cold spot is added in the design, and the system reduces the influence of the error of the Ideal gas law by bringing in the cold spot for compensation. The temperature of the cold spot varies from -20°C to 30°C. The following empirical equation is used to calculate the saturated Vapor pressure[7]:

$$\log(p/Pa) = 5.006 + A + BT^{-1} + C\log T + DT^{-3} \qquad (3.1)$$

The empirical equation can effectively calculate the Vapor pressure with an accuracy of ± 5% or higher. In the equation, A, B and C are constants for calculating the cesium Vapor pressure, and T is the absolute temperature, where A is 4.165, B is -3830, and there are no constants C and D. The empirical equation obtained by substituting constants is:

$$\log(p/Pa) = 9.171 - \frac{3830}{T} \qquad (3.2)$$

The Ideal gas law is a Equation of state that describes the relationship between pressure, volume, amount of substance and temperature when Ideal gas is in equilibrium. The expression is:

$$pV = nRT \qquad (3.3)$$

The number of particles is calculated as follows:

$$N = \frac{p \times V \times NA}{R \times T} \qquad (3.4)$$

Where *N* is the Particle number, *p* is the pressure, *V* is the volume, *NA* is the Avogadro constant, *T* is the absolute temperature, and *R* is the Gas constant.

According to the above saturated Vapor pressure theory and the Ideal gas state, it can be calculated that the concentration of cesium atoms in the closed cavity changes with temperature.

Figure 1 shows the structure of the cesium vapor pool. Cesium is a light golden yellow active metal that is solid at room temperature, has a low melting point, high reactivity with oxygen and water, and can form compounds with various elements. Therefore, stainless steel is chosen as the cesium cell material. The cesium cell consists of a cross shaped Vacuum chamber, with a length of 150.8mm and a volume of $7.70823 \times 10^{-5} m^3$. Cesium is stored at position A of the Vacuum chamber, and the cesium vapor is evaporated to the whole chamber by heating. A thermocouple is set at position B to detect the temperature of the cavity and facilitate the control of the amount of cesium vapor. Ruby window is selected at position C at both ends of the optical path in the pool, which can increase the transmittance of infrared laser. Position D is a detachable flange used to inject cesium into the cavity. The E position valve is used for Vacuum chamber vacuum pumping. The position F is for semiconductor cooling sheets for cold spot cooling.

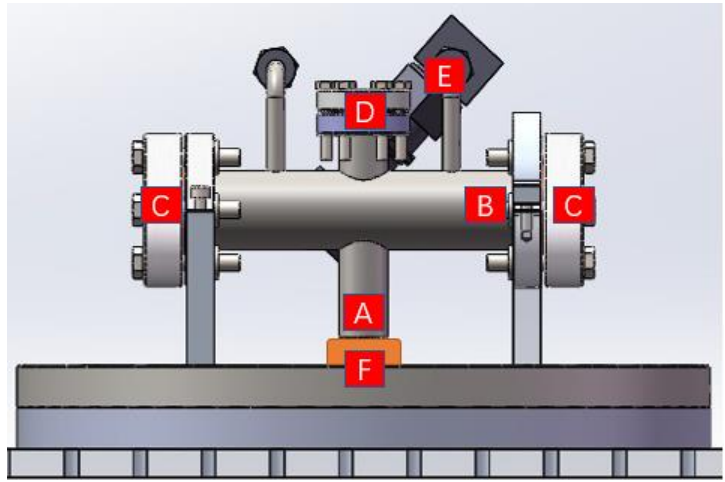

**Figure 1.** Structure diagram of cesium vapor cell

### 3.2 Temperature control system

In the experiment, to obtain cesium vapor, it is necessary to heat the cesium in the vapor cell to evaporate it into the cesium cell in the form of vapor[8]. This system uses the SPEC-TE2 intelligent temperature controller model from Chuangpu Instrument Technology Co., Ltd. (Part 4 in Figure 3) to control the amount of cesium vapor by wrapping heating wires on the outer wall of the cesium vapor cell. This temperature controller can set up to 60 temperature control segments as needed to heat up, maintain insulation, and cool down the cesium cell, and monitor the real-time temperature of the cesium cell through thermocouples.

## 4. Systems design

### 4.1 Experimental setup

The schematic diagram of the cesium atomic concentration detection system structure is shown in Figure 2, which mainly consists of three parts: a signal generation driving unit, an optical transmission testing unit, and a signal acquisition and processing unit.

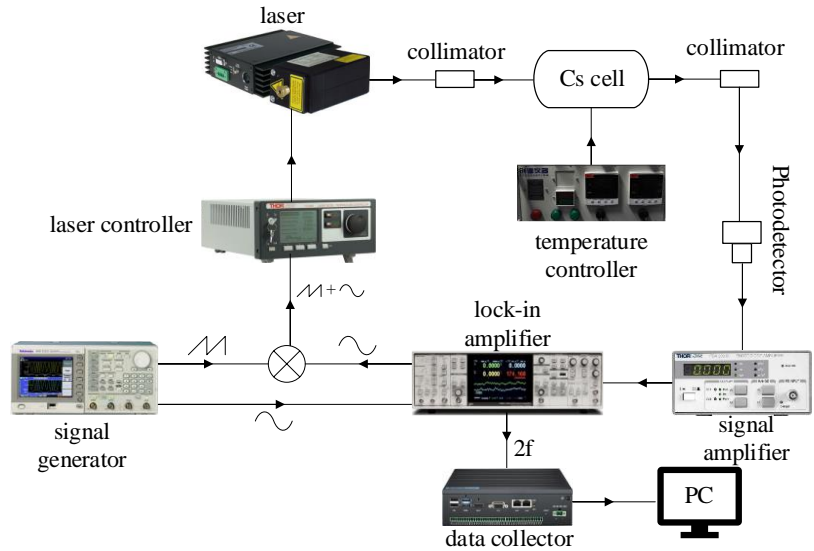

**Figure 2.** Schematic diagram of the cesium atom concentration detection system structure

    The signal generating unit includes a signal generator, a laser controller, and a laser, mainly used to generate the required wavelength of laser for experiments. The signal generator outputs the low-frequency Sawtooth wave scanning signal superimposed with the high-frequency sine wave signal, which is loaded onto the laser through the laser controller to achieve the scanning and modulation of the required laser wavelength. The optical transmission unit includes a collimator, a filter, and a controllable cesium vapor cell. After being collimated by the collimator, the modulated laser is absorbed and attenuated by the cesium vapor cell, which is then transmitted to the photodetector. The cesium vapor is obtained by heating the cesium cell through a temperature controller. The signal acquisition and processing system includes a photodetector, a signal amplifier, a Lock-in amplifier, a data acquisition card and signal analysis software. The photodetector converts the attenuated optical signal into an electrical signal, which is amplified by the signal amplifier and then transmitted to the Lock-in amplifier. The Lock-in amplifier filters the noise and extracts the second harmonic of the signal, Finally, the data acquisition card converts electrical signals into digital signals and transmits them to computer software for data processing. Figure 3 shows the overall diagram of the constructed experimental system.

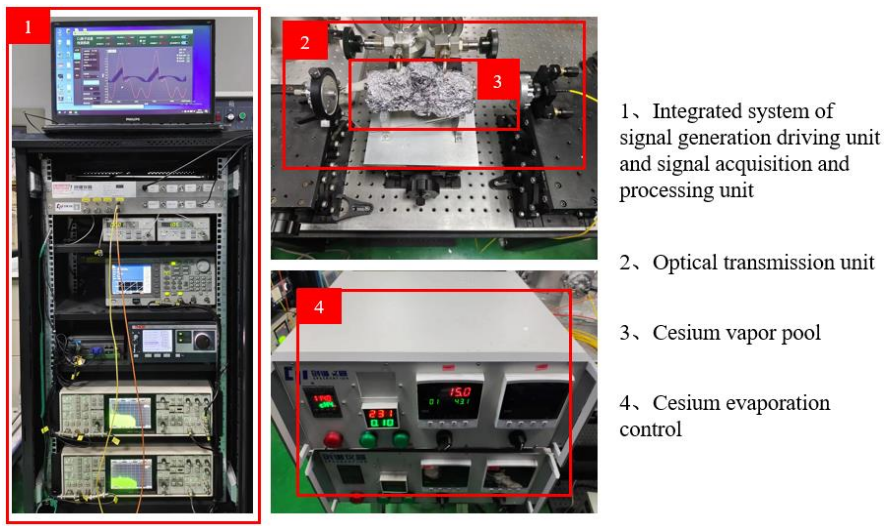

**Figure 3.** Experimental diagram of cesium atom concentration detection system

### 4.2 Qualitative testing of the system

Main experimental parameters: low-frequency Sawtooth wave signal frequency is 50Hz, Sawtooth wave amplitude is 1000mVpp, high-frequency sinusoidal modulation signal frequency is 4kHz, and sinusoidal output amplitude is 150mV; The driving current of the laser is 76 mA, and the working temperature of the laser is 35.7°C; The phase of the Lock-in amplifier is about -127.6°, the output voltage sensitivity is 10mV, and the time constant is 1ms.

By controlling the temperature of the cesium vapor cell for cesium vapor evaporation, the cesium vapor cell contains a certain concentration of cesium vapor[9]. The signal generator outputs low-frequency Sawtooth wave scanning signal superimposed with high-frequency sine wave signal, which is loaded onto the laser through the laser controller to realize scanning and modulation of 852.1nm laser wavelength. Build a system absorption optical path and collimate the modulated laser through a collimator before passing through the cesium vapor cell. The cesium vapor will attenuate the laser and generate an absorption signal, as shown in channel 1 of Figure 4. The photodetector converts the attenuated optical signal into an electrical signal, which is amplified by the signal amplifier and transmitted to the Lock-in amplifier. The Lock-in amplifier filters the noise and extracts the second harmonic from the signal to obtain the second harmonic signal proportional to the concentration, as shown in Figure 4.4, Channel 2. Based on the following experimental results, it can be determined that the system can qualitatively detect cesium.

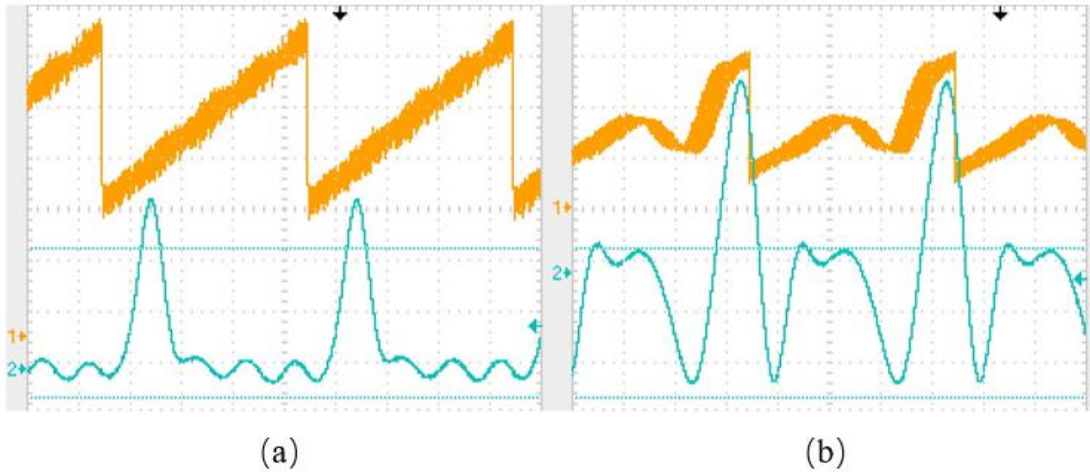

**Figure 4.** (a) Original signal map after modulation, (b) Absorption signal map

### 4.3 Quantitative detection of the system

Set the temperature of the temperature controller to gradually increase in a gradient trend, starting from 30 °C and increasing by 2 °C to 52 °C each time. Set a rise time of 10 minutes and a holding time of 30 minutes for each gradient to stabilize the cesium vapor in the steam cell[10]; By collecting the amplitude of the second harmonic signal to calibrate the cesium concentration, the trend of cesium concentration changing with the cesium cell temperature is shown in Figure 5.

The experimental results show that the concentration of cesium atoms increases with the increase of temperature, and the trend of the two is the same. This indicates that the cesium vapor cell can achieve temperature and concentration control, and the system can quantitatively detect the concentration of cesium in the cesium cell[11][12]. But when the temperature of the cesium cell exceeds 52 °C, that is, when the concentration in the cavity increases to a certain range, the concentration no longer changes, the absorption signal tends to saturation, and the absorption peak widens and flattens, as shown in Figure 6. Therefore, it can be concluded that the system detection range is within $5\times10^6$-$2.5\times10^7$N/cm$^3$, system resolution better than $1\times10^6$ N/cm$^3$.

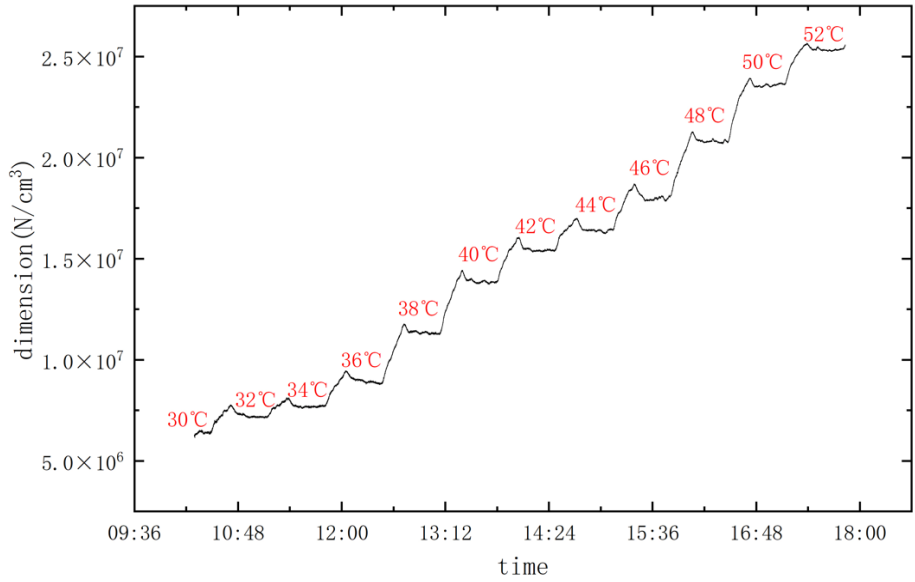

**Figure 5.** Trend chart of cesium concentration changing with cesium cell temperature.

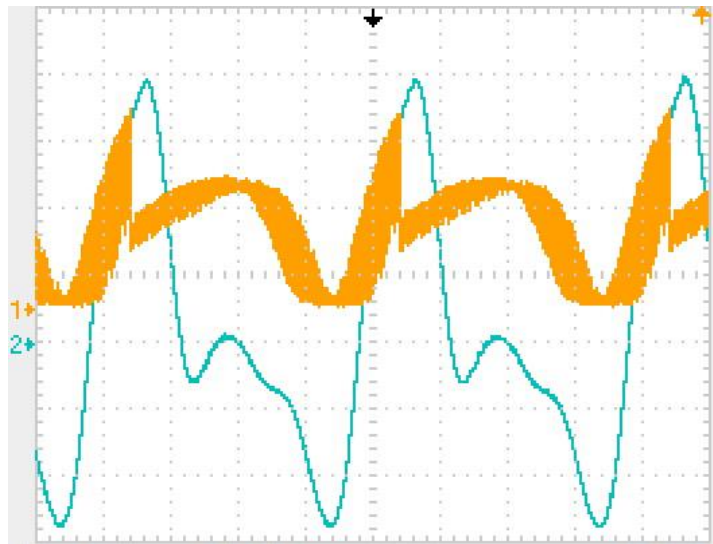

**Figure 6.** Signal under absorption saturation

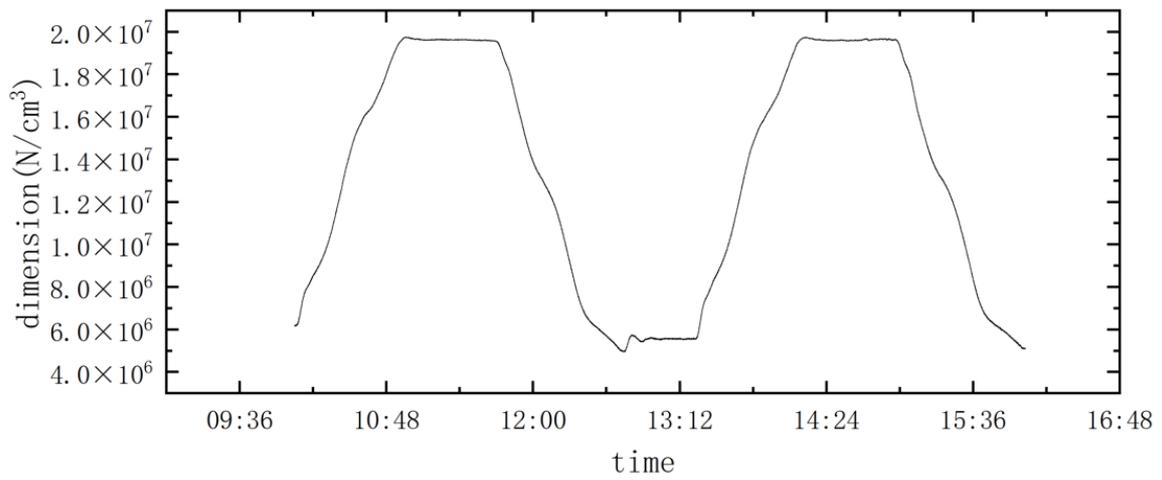

**Figure 7.** System Stability Experiment Results

### 4.4 Stability of the system

The main purpose of this experiment is to verify the stability of concentration (harmonic signal) changes under a continuous temperature change. Heat the cesium cell from 30 °C for 1 hour to 52 °C, keep it warm for 45 minutes, and then cool it down to 30 °C for 1 hour. Repeat the experiment twice continuously, and the results are shown in Figure 7. The trend of concentration increase and decrease in the two experiments is basically the same, and the error of the highest concentration in the two experiments is within 1.44%. From the experimental results, it can be concluded that the system has good stability.

### 5. Conclusion

This article introduces the design and functional testing of a cesium atom concentration detection system based on TDLAS technology, which detects the concentration and changes of cesium atoms using a cesium vapor pool. Through testing, it was found that the system can perform qualitative and quantitative detection of cesium. The gradient curve of concentration variation with temperature shows that the detection range of the system is within $5\times10^6$-$2.5\times10^7$ N/cm$^3$, system resolution better than $1\times10^6$ N/cm$^3$. Two consecutive repeated experiments were conducted on the system, and the trend of concentration change was basically the same. The maximum concentration error between the two experiments was within 1.44%, indicating that the system has good stability. The cesium atom concentration detection technology of this system has a guiding role in optimizing the performance of negative ion sources.

### Acknowledgments

This work was supported by the HFIPS Director's Fund (YZJJQY202204 and 2021YZGH02), Comprehensive Research Facility for Fusion Technology Program of China under Contract No. 2018-000052-73-01-001228 and National Key R&D Program of China (2017YFE300103, 2017YFE300503)### References